# MACRO AND MICRO SCALE ELECTROMAGNETIC KINETIC ENERGY HARVESTING GENERATORS

S. P. Beeby[1], M. J. Tudor[1], R. N. Torah[1], E. Koukharenko[1], S. Roberts[2], T. O'Donnell[3], S. Roy[3]

[1] University of Southampton, School of Electronics and Computer Science, Southampton, UK.
[2] Perpetuum Ltd, www.perpetuum.co.uk
[3] Tyndall National Institute, Prospect Row, Cork, Ireland.

## ABSTRACT

## 1. INTRODUCTION

Developments in ultra low power electronics, RF communications and MEMS sensors in wireless sensor networks has led to the requirement for in-situ power supplies capable of harvesting energy from the environment. This paper is concerned with generators that harvest electrical energy from the kinetic energy present in the sensor nodes environment. These generators have the potential to replace or augment battery power which has a limited lifetime and requires periodic replacement which limits the placement and application of the sensor node.

This paper presents the standard analysis of the energy available in vibration energy generators and discusses the practical implications of reducing electromagnetic generators down in size to the micro level. The practical results from micro devices fabricated under an EU project entitled Vibration Energy Scavenging (VIBES) will be presented. The implications for further reducing the generator in size and employing fully integrated micromachining processes will also be discussed.

## 2. BASIC THEORY

Resonant generators are essentially a second-order, spring-mass system. Figure 1 shows a general example of such a system based on a seismic mass, *m*, on a spring of stiffness, *k*. Energy losses within the system are represented by the damping coefficient, $c_T$. These losses comprise parasitic losses $c_p$ (e.g. air damping) and electrical energy extracted by the transduction mechanism, $c_e$. These generators are intended to operate at their natural frequency, given by $\omega_n=(k/m)^{1/2}$, and should be designed such that this coincides with the vibrations present in the intended application environment. The theory of inertial-based generators is well documented [1,2] and will only be briefly covered here. Assuming the generator is driven by an external sinusoidal vibration of the form $y(t)=Y\sin(\omega t)$, it will move out of phase with the mass at resonance resulting in a net displacement, $z(t)$, between the mass and the frame.

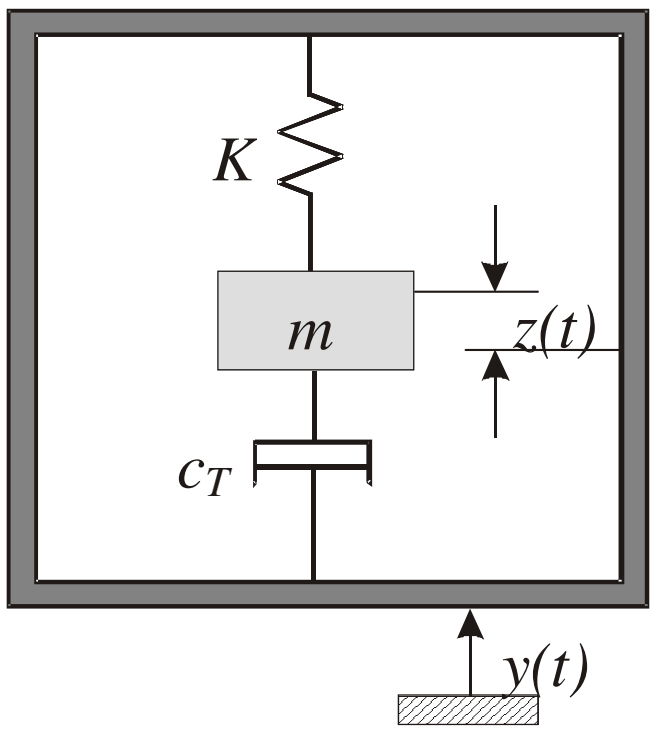

Figure 1 - Model of a linear, inertial generator.

Assuming that the mass of the vibration source is significantly greater than that of the seismic mass and therefore not affected by its presence, then the differential equation of motion is described as:

$$m\ddot{z}(t)+c\dot{z}(t)+kz(t)=-m\ddot{y}(t) \qquad (1)$$

Since energy is extracted from relative movement between the mass and the inertial frame, the following equations apply.

The standard steady state solution for the mass displacement is given by:

$$z(t)=\frac{\omega^2}{\sqrt{\left(\frac{k}{m}-\omega^2\right)^2+\left(\frac{c_T\omega}{m}\right)^2}}Y\sin(\omega t-\phi) \tag{2}$$

Where $c_T$ is the total damping factor and $\phi$ is the phase angle given by:

$$\phi=\tan^{-1}\left(\frac{c_T\omega}{(k-\omega^2 m)}\right) \tag{3}$$

The power dissipated within the damper (i.e. extracted by the transduction mechanism and parasitic damping mechanisms) is given by:

$$P_d=\frac{m\zeta_T Y^2\left(\frac{\omega}{\omega_n}\right)^3\omega^3}{\left[1-\left(\frac{\omega}{\omega_n}\right)^2\right]^2+\left[2\zeta_T\left(\frac{\omega}{\omega_n}\right)\right]^2} \tag{4}$$

where $\zeta_T$ is the total damping ratio given by $\zeta_T=c_T/2m\omega_n$. Maximum power occurs when the device is operated at $\omega_n$ and in this case the theoretical maximum power stored in the system is given by:

$$P=\frac{mY^2\omega_n^3}{4\zeta_T} \tag{6}$$

Equation 6 demonstrates that the power delivered varies linearly with the mass. Also, for a constant excitation amplitude, power increases with the cube of the frequency, and for a constant frequency power increases with the square of the base amplitude. It should be noted that since acceleration $A=\omega^2 Y_0$ both of these conditions must be associated with increasing acceleration in the environmental vibrations. Incorporating the parasitic and electrical damping gives the power delivered to the electrical load:

$$P_L=\frac{m\zeta_e Y^2\omega_n^3}{4(\zeta_p+\zeta_e)^2} \tag{7}$$

Maximum power is delivered when $\zeta_p=\zeta_e$. The damping factor arising from electromagnetic transduction $c_e$ can be estimated from equation 8.

$$c_e=\frac{(NlB)^2}{R_L+R_{coil}+j\omega L_{coil}} \tag{8}$$

where $N$ is the number of turns in the generator coil, $l$ is the side length of the coil (assumed square), and $B$ is the flux density to which it is subjected and $R_L$, $R_{coil}$, and $L_{coil}$ are the load resistance, coil resistance and coil inductance respectively. Equation 8 is an approximation and only ideal for the case where the coil moves from a high field region B, to a zero field region. Equation 8 demonstrates that $R_L$ can be used to adjust $c_e$ to match $c_p$ and therefore maximise power but this must be done with the coil parameters in mind. It can be shown that the optimum $R_L$ can be found from equation 9 and the maximum average power delivered to the load can be found from equation 10 [3].

$$R_L=R_{coil}+\frac{(NlB)^2}{c_p} \tag{9}$$

$$P_{L\max}=\frac{m\omega_n^3Y^2}{16\zeta_p}\left(1-\frac{R_{coil}}{R_{load}}\right) \tag{10}$$

## 2. ELECTROMAGNETIC GENERATOR CONFIGURATION

The generators presented in this paper are based upon a cantilever beam structure with a magnetic circuit comprising four Neodymium Iron Boron (NdFeB) magnets and a traditionally wound coil [4]. The magnets are located either side of the coil and the magnetic circuit is completed by the mild steel keepers shown in figure 2. Either the coil can move relative to the magnets or the magnets move relative to the coil. NdFeB magnets were selected since they exhibit the greatest energy density therefore maximising the flux gradient that cuts the coil.

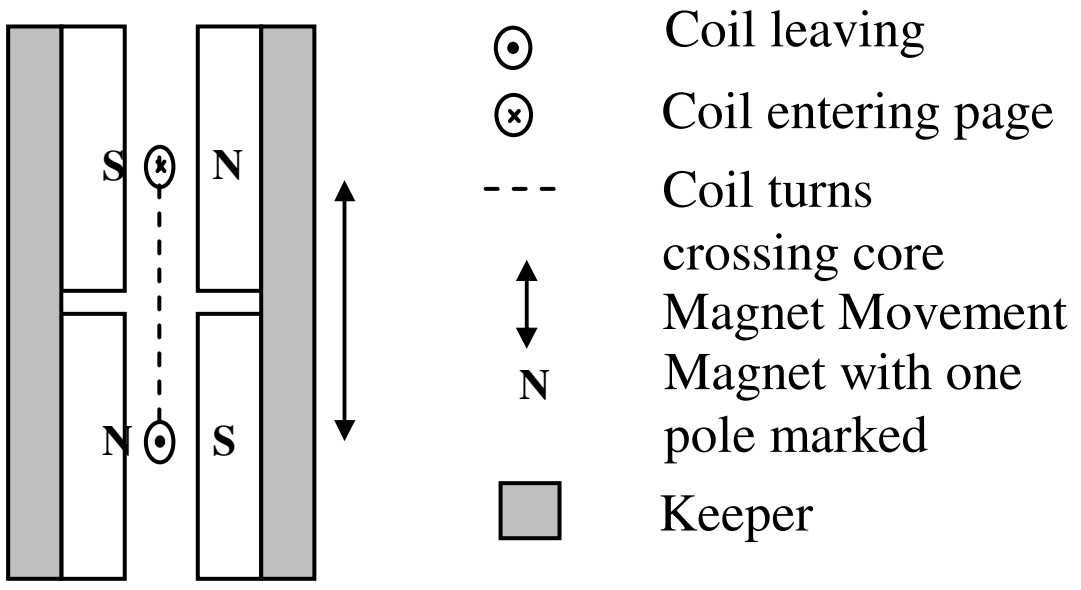

Figure 2 – Cross section through the four magnet, keeper and coil arrangement

## 3. MACRO AND MICRO GENERATORS

The macro generator, Perpetuum PMG7 is shown in figure 3. This is designed to resonate at 50Hz and produces about 3mW AC power at 0.5m/s$^2$ RMS acceleration. The active mass of the generator is 85g and its volume is 41.3cm$^3$.

Two electromagnetic microgenerators have been developed during the VIBES project based upon the configuration of figure 2. Both of these generators use discrete coils and magnets combined with micro fabricated components. The first is a laterally vibrating silicon structure shown in figure 4 which has a traditionally wound copper coil inserted into the moving proof mass. The coil moves in the plane of the chip relative to the fixed magnets.

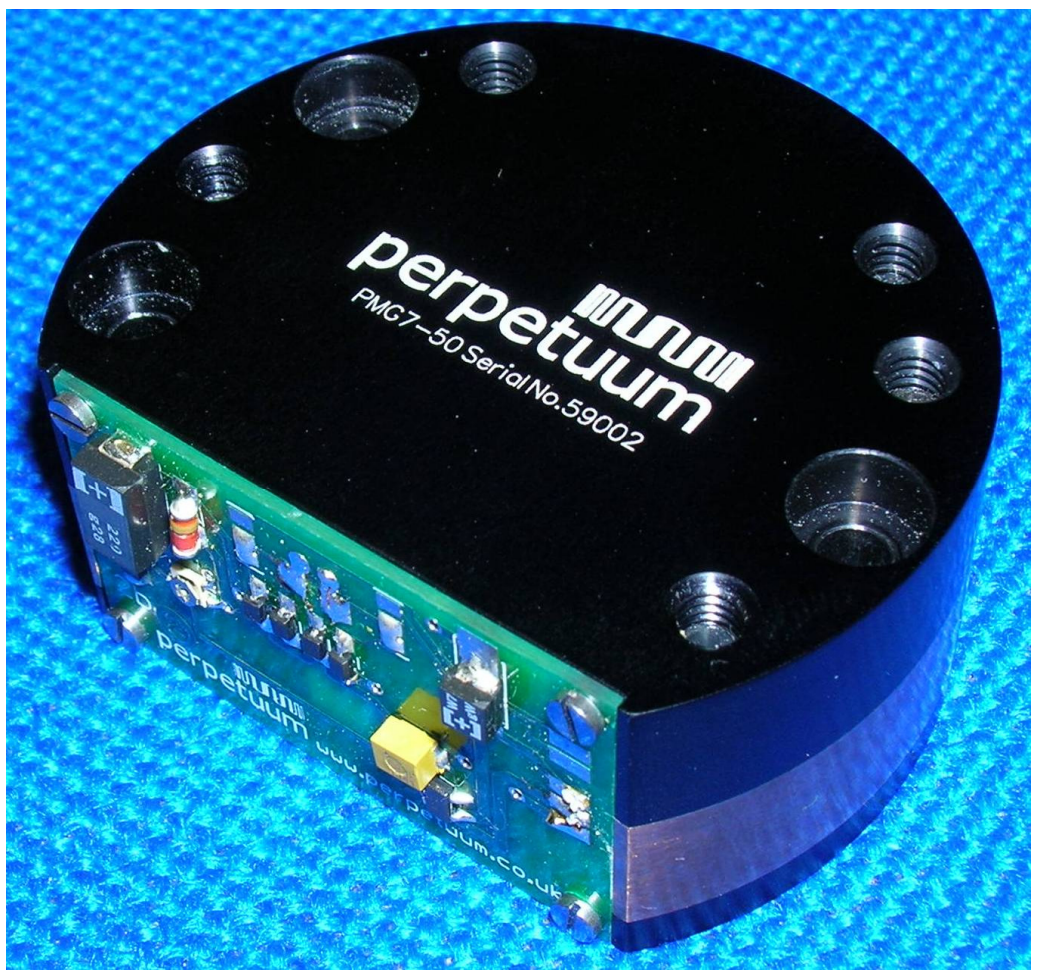

Figure 3 – Perpetuum PMG7 generator

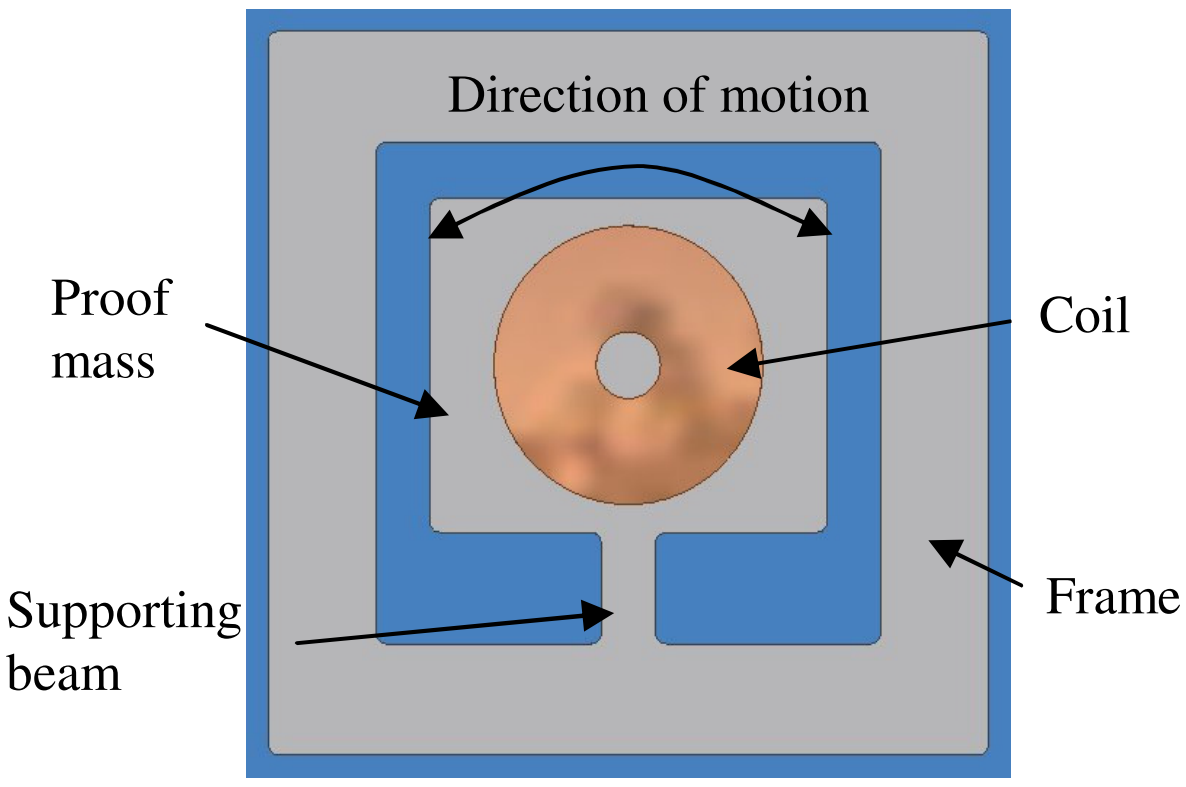

Figure 4 – Laterally vibrating silicon microgenerator

The silicon structure is formed by deep reactive ion etching through the silicon wafer thickness. The natural frequency of the structure is controlled by the width, length and thickness of the supporting beam and the mass of the paddle. The beam is 1mm long, 0.5mm wide and 525µm thick. The resulting total proof mass is 0.028 grams. The copper coil has an outer diameter of 2.4mm, inner diameter of 0.6mm and contains 600 turns. The coils were wound from an enamelled copper wire with a diameter of 25µm. The maximum deflection of the paddle is limited by the chip frame and this equates to a coil displacement of 240µm. Electrical connection is made by taking the coil wires out along the top surface of the cantilever. In practice this is very problematic since it involves adhesively bonding the wires along the cantilever. This has led to increased mechanical damping and a wide variation in the performance of these devices. Full details of the design and fabrication of this generator can be found in reference 5. The resonant frequency of this device is 9.5 kHz and it produces 122nW into a 110 Ohm load at 3.5m/s$^2$ acceleration. The volume of this generator is 68mm$^3$.

The second microgenerator was developed in order to avoid the difficulty in making electrical connection to the coil encountered with the laterally vibrating device. The revised generator is shown in figure 5. Since this has not been presented previously this device is described in more detail.

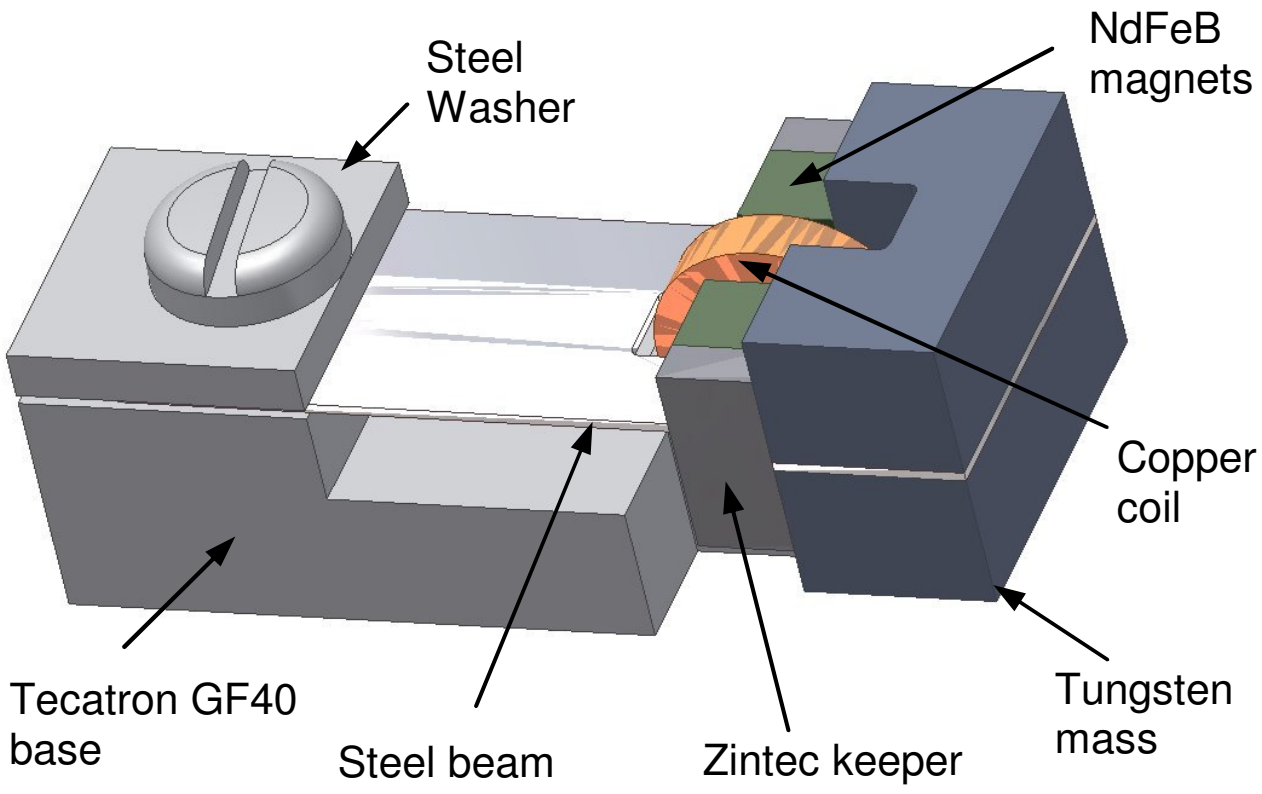

Figure 5 – Micro cantilever generator

In this configuration the 4 magnets are located at the free end of a cantilever beam. This beam is clamped at one end between a plastic base and steel washer by an M1 nut and bolt. The magnets move vertically relative to the coil that is adhesively bonded to the plastic base. Due to the cyclical stressing of the beam during operation, single crystal silicon, stainless steel and beryllium copper were chosen as suitable beam materials. Single crystal silicon is elastic to fracture and therefore will not change its material properties or dimensions as a result of being cyclically stressed. Silicon also does not suffer from fatigue failure. Silicon beams can be fabricated by DRIE etching through the thickness of a wafer. Stainless steel type 302FH also has excellent elastic properties and fatigue characteristics without the brittleness of silicon. Stainless steel and beryllium copper beams can be fabricated by wet chemical etching through sheet metal although the dimensional tolerances are not as tight as for silicon etching.

As stated previously, the available mechanical energy in the system can be increased by maximising the inertial mass at the free end of the beam. In the cantilever arrangement the mass of the magnets is included in the inertial mass of the system and there is also space available on the beam end for additional mass. For this purpose, machinable tungsten elements were formed by wire erosion. Machinable tungsten was chosen for its high density of ~19000kgm$^{-3}$. The density of the NdFeB magnets is ~7600kgm$^{-3}$. The magnetic keepers were wire eroded from Zinc coated mild steel (Zintec), which has high permeability but maintains a high corrosion resistance due to the zinc coating. The total proof mass for this assembly is 0.44 grams. The volume of the active components including the free space required for the proof mass to oscillate is 60mm$^3$. The components were assembled onto a base milled from Tecatron GF40 plastic. The finished generator was subsequently mounted on an 8 pin DIL package for attachment to a shaker.

The beam thickness has been defined by finite element analysis, the model of the cantilever and attached components being shown in figure 6.

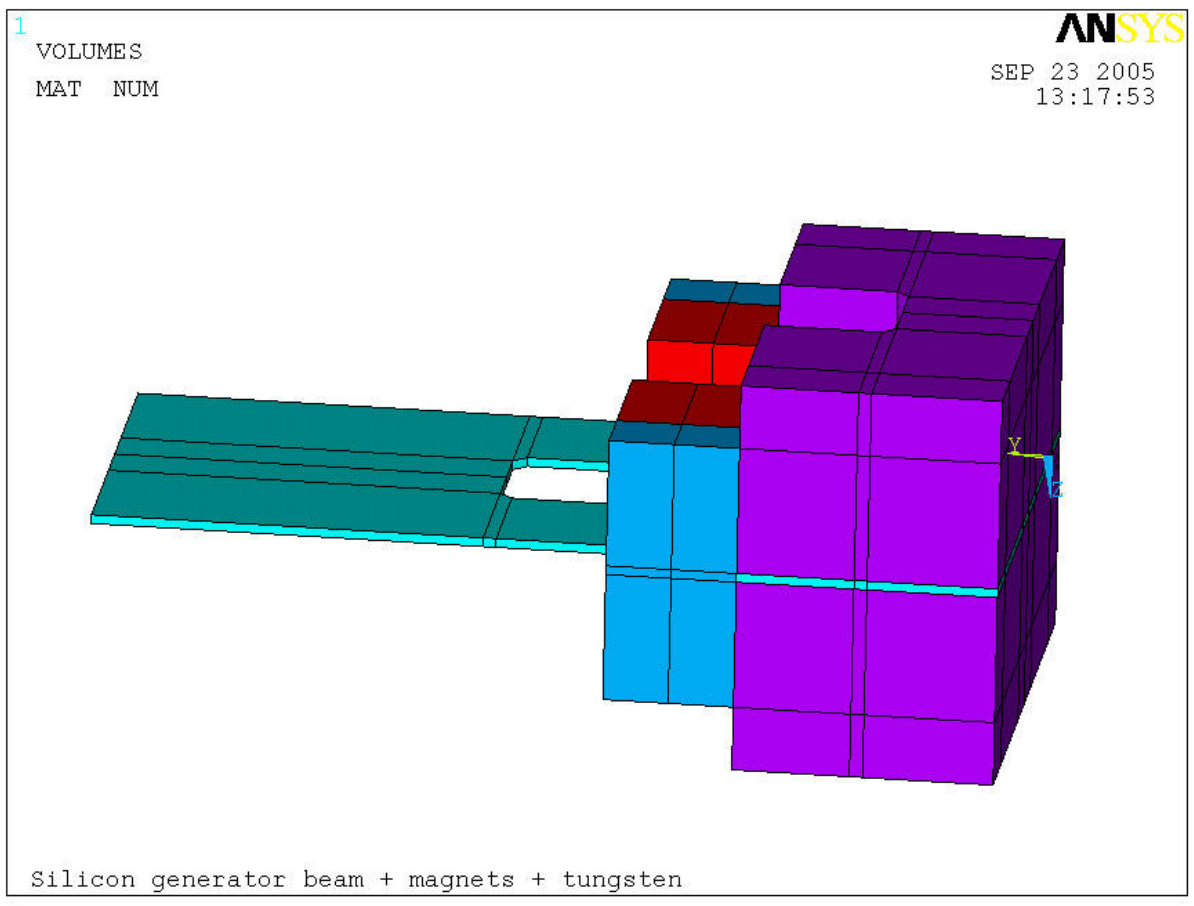

Figure 6 – FE model of cantilever and components

The simulated resonant frequencies for a range of beam thicknesses and materials are given in figure 7. These are based upon standard sheet metal thicknesses and bespoke wafers are required for the silicon case.

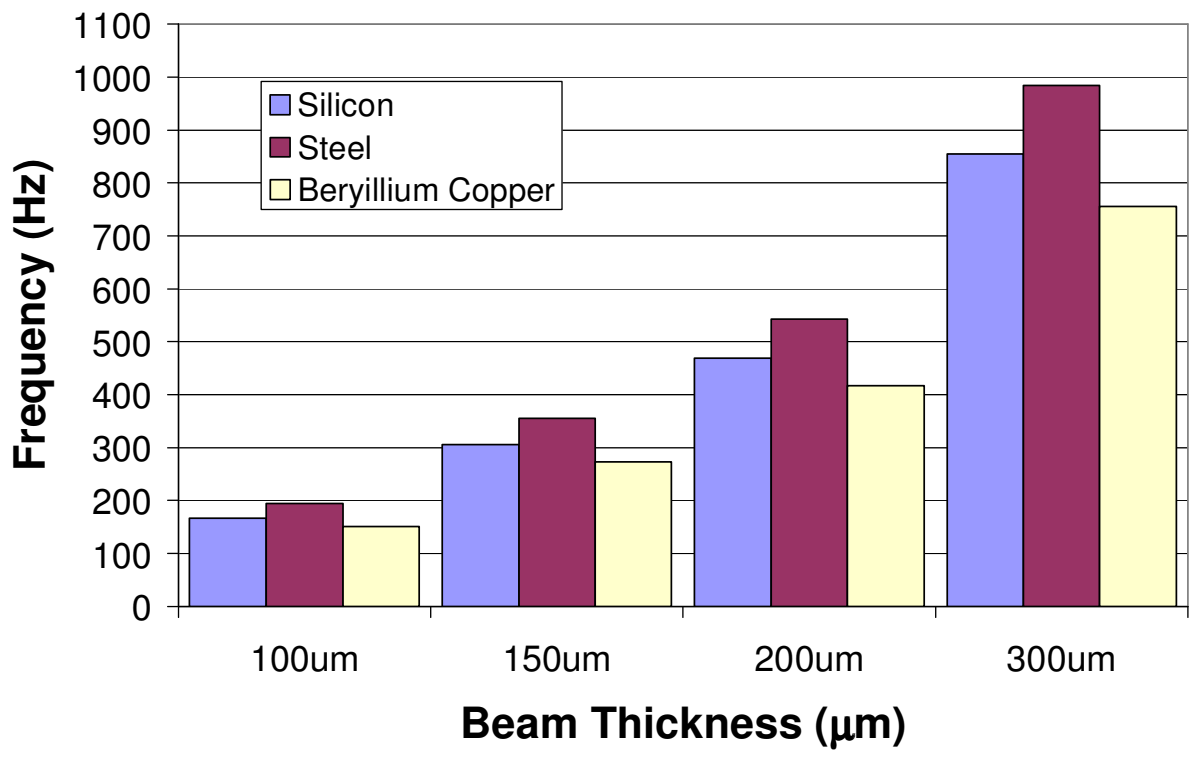

Figure 7 - Generator frequency for varying beam thickness and material

A generator with a 150µm thick stainless steel beam has been assembled and the initial results are as follows. The open circuit output of the generator has been plotted in order to calculate the parasitic damping associated with the device. The Quality factor can be calculated from figure 8 and was found to be 216.

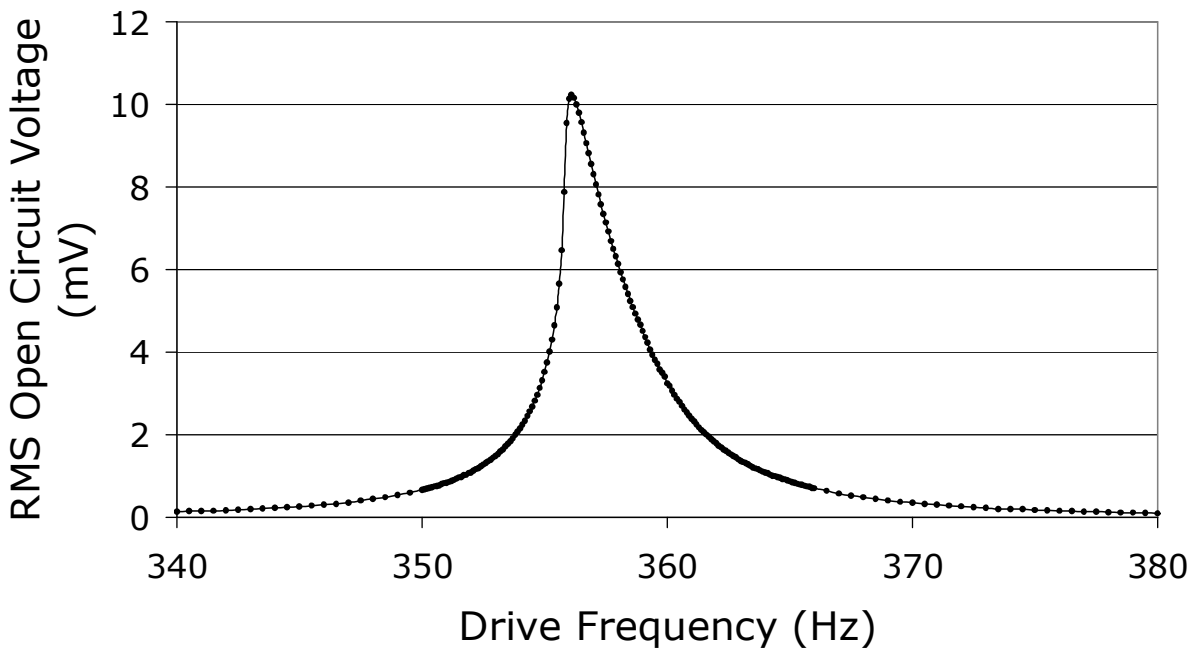

Figure 8 - RMS Open Circuit Voltage versus Drive Frequency at 1m/s$^2_{RMS}$

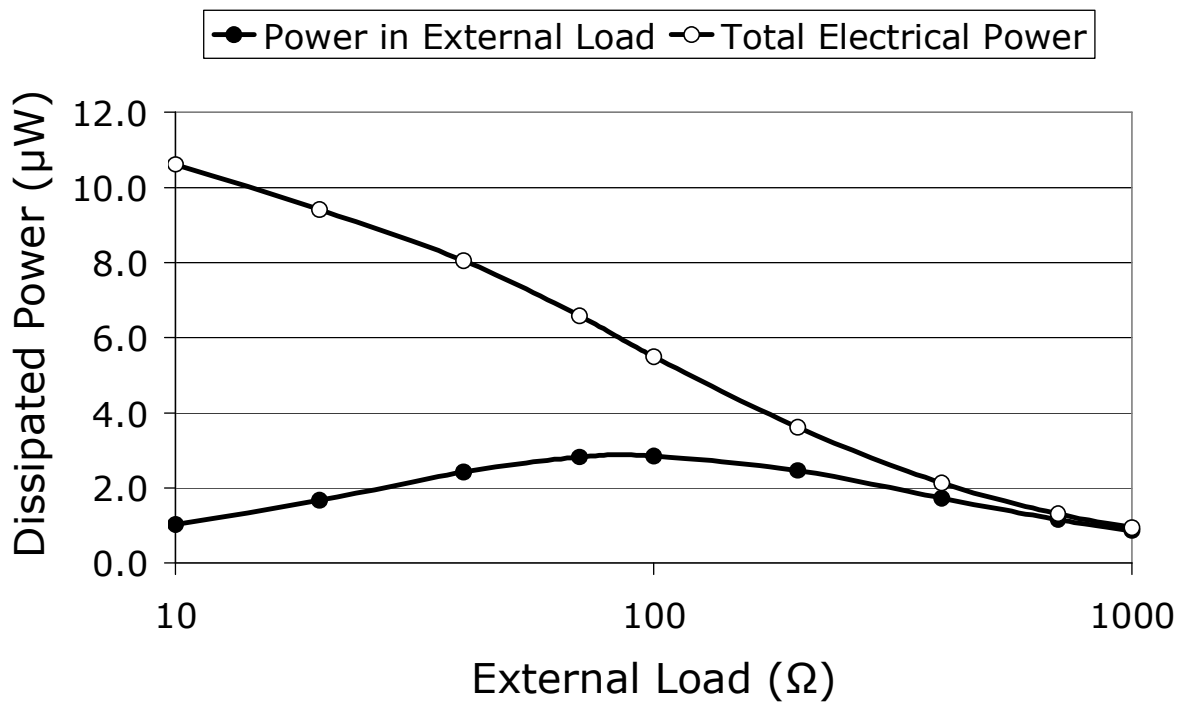

Figure 9 - Power in External load and Total Electrical Power a=3m/s$^2_{RMS}$ and f = 350Hz

The maximum power delivered to the load from an excitation acceleration of 3m/s$^2$ was 2.85µW at 350Hz,

with a resistive load of 100 Ohms as shown in figure 9. The total power is that dissipated in the coil plus the power delivered to the load.

## 4. DISCUSSION OF RESULTS

The generators described here can be simply compared by calculating their power densities in terms of $nW/mm^3$. In order to make this a valid comparison, the power outputs have been normalised to the acceleration used in the testing of the micro cantilever device which is $3m/s^2$. Given this acceleration, the macro generator demonstrates a power density of $2615nW/mm^3$. The laterally vibrating microgenerator exhibits a power density of $1.3nW/mm^3$ and the cantilever microgenerator $47nW/mm^3$.

The macro generator exhibits the highest power density due to its use of mm scale magnets providing increased flux density and the generator exhibits a greater magnet velocity even at $0.5m/s^2$ excitation acceleration. Both micro devices suffer in comparison with the macro generator due to reduced coupling between the coil and magnets. This is highlighted by the electrical damping coefficient which can be calculated for the cantilever microgenerator from the Q-factor of the device under different load conditions as follows. With a 100 Ohm load the overall Q of the device, $Q_T$, which includes both electrical and parasitic damping equals 181. This compares to a $Q_{OC}$ (parasitic damping only) of 216 and, given $Q_{oc} = 1/(2\zeta_p)$, $\zeta_P$ equals 0.0023. The Q factor of the generator with parasitic damping ignored can be calculated from equation 11.

$$\frac{1}{Q_T} = \frac{1}{Q_{OC}} + \frac{1}{Q_E} \qquad (11)$$

This gives a $Q_E$ of 1120 which equates to a $\zeta_e$ of 0.00045. It is clear the condition for optimum power delivery, which occurs when $\zeta_p = \zeta_e$, has not been met. This is confirmed by the fact the optimum load resistance matches the coil resistance (~93 Ohms) which occurs when $\zeta_p >> \zeta_e$ [6].

In order to increase the electromagnetic damping factor, and therefore improve the electromagnetic coupling, the variables given in equation 8 should be optimised. For example, increased $\zeta_e$ could be achieved by increasing the magnetic field and the number of coil turns within the given volume and optimising $R_L$. For the devices presented here, the flux densities for the lateral micro generator is 0.29T and for the cantilever microgenerator 0.41T. This could be improved by reducing the gap between the magnets, something which is feasible in both cases. Furthermore, the flux linkage (defined as the amount of flux cutting the coil) can be improved by increasing the magnet size.

The reduced power density of the two microgenerators is also due to the limited amplitudes of vibration of the proof masses during the vibration cycle. The amplitude of the masses, $z_0$, in the microgenerators can be calculated from $z_t = QY_t$. The lateral generator has a loaded Q factor of 164 and $Y_0$ equals 1nm, hence the mass/coil displacement is just 164nm. The cantilever micro generator has a loaded Q of 350 and a base amplitude of 0.62μm giving a $z_0$ of 217μm. Increasing the amplitude for a given frequency will increase the relative velocity between the magnets and coil and therefore induce a greater voltage in the coil. The implications of increasing this amplitude are discussed below.

Alternatively, $\zeta_p = \zeta_e$ could be achieved by reducing parasitic damping so that $Q_{OC}$ increases. This would also increase the velocity of the inertial mass. However, high generator Q factors could lead to unacceptably large inertial mass amplitudes. The maximum amplitude will be limited by the maximum cyclical stress the generator material can withstand. Given this consideration, single crystal silicon is attractive since it is elastic to fracture and possesses a very high intrinsic Q. The maximum stress will in practice depend upon crystalline imperfections and surface irregularities. This highlights an important difference between the lateral and cantilever micro generators. In the case of the lateral generator, the etched side walls of the supporting beam experience the maximum stress as the mass deflects and this is a relatively rough surface with many imperfections. The practical strength of this structure is therefore much less than the theoretical maximum strength of silicon. The cantilever microgenerator is designed to vibrate out of the plane of the wafer and the beam can be etched from double polished silicon wafers which provide the optimum surface finish conditions to reduce the risk of beam fracturing. The practical strength will vary from beam to beam and it is therefore sensible to operate the generators at amplitudes which provide sufficient margin of safety to ensure the fracture stress is not exceeded. Nonetheless, this configuration is capable of much higher amplitudes of vibration than the lateral case. The maximum amplitudes of the stainless steel and beryllium copper beams will be limited by the fatigue characteristics of the material.

This analysis has concentrated on power densities but another practical limitation of reduced generator size is the voltage generated. Whilst the present cantilever microgenerator delivers a useful amount of power to the load (2.85µW), the voltage is only 16.9mV. This is too low to simply rectify with a diode and requires stepping up to useful levels. Measures to increase the delivered power will naturally benefit the generated voltage but in practice parameters may need to be optimised for maximum voltage rather than maximum power generation.

## 5. CONCLUSIONS

This paper has presented a practical comparison between a macro electromagnetic generator and two microgenerators based upon the same magnetic circuit. The results demonstrate the difficulty in achieving sufficient electromagnetic coupling in order to approach the theoretical levels of power and voltages as the generators reduce in size. In order to achieve the $\zeta_p=\zeta_e$ condition for the generators presented, coil turns should be maximised within the given volume and magnet size and spacing should be selected to maximise flux linkage and the generator. As a general rule, the generators should be operated at maximum inertial mass displacement and should be designed such that this value is as large as possible. Of course, this can only be done with exact knowledge of the excitation vibrations present in the intended application as is fundamentally the case for all vibration based generators. Further reductions in electromagnetic generator size, in particular the integration of micromachined structures with electroplated coils and magnets rather than discrete components, may not be able to deliver practical voltage and power levels. Piezoelectric transduction that inherently produces higher voltage levels may be better suited to smaller micromachined inertial based generators.